# Observation of well-defined quasi-particle over a wide energy range in quasi-2D system

**Authors**: L.-L. Jia[1], Z.-H. Liu[1], Y.-P. Cai[1], T. Qian[2], X.-P. Wang[2], H. Miao[2], P. Richard[2,3], Y.-G. Zhao[1], Y. Li[1], , M. Shi[4], H. Ding[2,3], D.-M.Wang[1], J.-B. He[1], G.-F. Chen[1,2], S.-C. Wang[1]*

[1] Department of Physics, Renmin University of China, Beijing, P.R. China 100872

[2] Beijing National Laboratory for Condensed Matter Physics, and Institute of Physics, Chinese Academy of Sciences, Beijing 100190, China

[3] Collaborative Innovation Center of Quantum Matter, Beijing, China

[4] Swiss Light Source, Paul Scherrer Institut, CH-5232 Villigen, Switzerland

**Abstract**

**We report the observation by angle-resolved photoemission spectroscopy measurements of a highly anisotropic Dirac-cone structure in high quality SrMnBi$_2$ crystals. We reveal a well-defined sharp quasi-particle, linearly dispersive with $v_{F_1}/v_{F_2} \sim 5-6$, forming a hole-like anisotropic Dirac-cone. The density of states for the cone remains linear up to as high as ~650 meV of binding energy. The scattering rate of the quasi-particle (QP) increases linearly as function of binding energy, indicating a non-Fermi-Liquid behavior. Our results suggest the existence of a dilute two-dimensional electron gas system in this three-dimensional material.**

**INTRODUCTION**

The description of condensed matter systems in the many-body limit is simplified by using the concept of quasi-particles, which represent elementary fermionic excitations behaving like particles[1,2,3]. In a Fermi liquid (FL), either three-dimensional (3D) or bi-dimensional (2D), these quasi-particles have a finite lifetime that decreases quadratically with energy[4,5]. Despite the recent enthusiasm generated by the recent discoveries of a plethora of new materials exhibiting Dirac cones and the relate anomalous properties[6,7,8,9], it is still not clear how the concept of quasi-particle can be extended to low-dimensional systems characterized by Dirac physics. Recently, Dirac-particle behavior with linear energy dispersion was predicted and observed in SrMnBi$_2$ and CaMnBi$_2$[10,11]. Indeed, large magnetoresistance (MR), high mobility and small cyclotron resonant mass suggest the existence of Dirac fermions in these materials[12,13]. Unlike the nearly symmetric Dirac cones in graphene though, angle-resolved photoemission spectroscopy (ARPES) measurements in SrMnBi$_2$[11] reveal a strong anisotropy in the Dirac cone dispersion, which LDA calculations attribute to the Bi $6p$ orbitals in the SrBi layer[11,14,15]. Both ARPES and thermopower measurements have reported the existence of small FS with linear dispersion, but are conflicting on the sign of carriers[11,14]. Quantum oscillation experiments also suggest a small Fermi surface (FS) compatible with Dirac points in the vicinity of the Fermi energy ($E_F$) [10].

Here, we report a detailed ARPES study on SrMnBi$_2$. A Dirac cone is observed with highly anisotropic Fermi velocities ($v_{F_1}/v_{F_2} \approx 5-6$). The linear dispersion of the Dirac cone is characterized by a sharp quasi-particle peak with scattering rate varying linearly with energy and a density of states (DOS) nearly constant over a wide energy range. Such behavior is inconsistent with the 3D and 2D FL description and suggests an unconventional 2D electron gas system in the bulk of this system.

The crystal structure of SrMnBi$_2$, as shown in Fig. 1a, consists of alternating SrBi and MnBi layers, with the unit cell doubling along the $c$-axis due to the opposite orientation of the MnBi$_4$ tetrahedra. The Bi atoms in the SrBi layer, separated by a distance of 3.239 Å, form a square net believed to be responsible for the interesting electronic states near the Fermi energy[14]. Well-defined peaks in the photoemission core level spectrum displayed in Fig. 1b can be assigned easily in the 3--30 eV range of binding energy. We note that the clear asymmetry in the Bi $5d$ peaks comes from the existence of two inequivalent Bi sites (Bi1 and Bi2 as shown in Fig. 1a) in the system. The more dispersive electronic states close to $E_F$ are mainly from the Bi $6p_{\frac{1}{2}}$ and the Mn $3d$ orbitals[14]. Figure 1c shows the integrated intensity within ±20 meV of $E_F$, which approximately represents the FS of the material. Two kinds of FSs can be distinguished: a large diamond-shape FS centered at the Brillouin zone center ($\Gamma$) originating from two bands mainly attributed to the Mn $3d$ orbitals[14], and four small crescent-like FSs located along $\Gamma$−M ((0, 0)--($\pi, \pi$)) and equivalent directions, with their long side perpendicular to $\Gamma$-M. The small FSs occupy 6.3% of the BZ area. The observed FS topology is consistent with quantum oscillation measurements and LDA calculations in the antiferromagnetic configuration[12,16].

We display in Fig. 2 the electronic band dispersion along high symmetry lines. The ARPES intensity plot, the corresponding second momentum derivative, and the momentum distribution curves (MDCs) along cut#1, perpendicular to the crescent FS (along Γ M), are shown in Figs. 2a-2c, respectively. We can identify two bands crossing $E_F$ near the BZ center. These two bands are mainly from the Mn 3$d$ orbitals located at the MnBi layer, forming the diamond-like FSs shown in Fig. 1c. The two bands are nearly coincident at $E_F$, as discussed later, and the corresponding FSs are hardly distinguished from the intensity plot in Fig. 1. Slightly away from the zone center, two linear bands dispersing in opposite directions converge near the $E_F$ at $k$=(0.37, 0.37)$\pi/a$. More precisely, linear extrapolation indicates that the apex of this Dirac cone electronic dispersion locates 10 meV above $E_F$. The inner band has lower intensity and is detected from $E_F$ to about $E_B \sim$ 100 meV. The outer band has higher intensity and a linear dispersion up to about 650 meV of binding energy with a well-defined sharp peak. At binding energy higher than 650 meV, the dispersion starts deviating from a linear dispersion and the peak becomes broader due to the proximity of other bands, as shown in Fig. 3a.

Figures 2d-2f show respectively the intensity plot, the second momentum derivative of intensity and the MDCs along cut#2. Similarly to cut#1, two dispersion branches with sharp peaks form a Dirac cone with apex slightly above $E_F$. However, the Fermi velocity $v_F$ is much smaller along that direction. While we find $v_F$ =8.64eV · Å (1.3×10$^6$ m/s) along cut#1, which corresponds to ~ 1/230 of the speed of light, similar to that of graphene, we determine a velocity about 5 times slower along cut#2 (1.78 64eV · Å $\approx \frac{1}{1100}$c). The Dirac cone is thus highly anisotropic. Although previously reported existence of an anisotropic Dirac cone in SrMnBi$_2$[11], the FS topology and carrier type are different between our results and that previous report.

In order to study in more detail the sharp dispersive feature associated with the Dirac cone, we recorded additional data along cut#1, using higher energy and momentum resolutions. Figure 3a shows the corresponding intensity plot. Its second momentum derivative intensity plot, displayed in Fig. 3d, enhances the bulk bands. A linear dispersion is observed over a wide range of energy, practically until the band reaches the bulk bands at high energy. To characterize the intrinsic properties of this dispersive feature, we removed a constant energy distribution curve (EDC) recorded away from the dispersive band assumed as a momentum-independent background. The result, illustrated in Fig. 3b for the low-energy part, shows a quite flat background with nearly zero intensity away from the band, as also illustrated by the corresponding background-removed EDCs shown in Fig. 3e, thus justifying our approach. We note that we observe that the peak splits by up to $0.016\text{Å}^{-1}$ close to $E_F$, which is not predicted by LDA calculations[11,14]. Although the origin of this splitting remains unclear, we conjecture that it may be caused by bi-layer interactions due to the doubling of the unit cell along the *c*-axis or to the spin-orbital coupling normally observed in heavy metals.

One of our most interesting observations is the nearly energy-independent quasi-particle spectral lineshape down to about 650 meV of binding energy. Indeed, Fig. 3c indicates that the momentum-integrated spectral intensity $n_\theta(\omega) \equiv \int_{k=k(\theta)} n(k,\omega) dk$ remains almost constant over a wide energy range, which is consistent with a 2D electron system. The total density of states in a 2D system with linear dispersion $\omega = \omega_0 + v_0(k-k_f)$ is linear as a function of binding energy

$$\rho(\omega) = \frac{2\pi}{v_0^2}(\omega - \omega_0) \tag{1}$$

Noticing that the size of the enclosed constant energy contour also increases linearly with binding energy for a Dirac-cone like dispersion, we thus estimate that the total DOS of the band increases linearly as a function of binding energy, as predicted by Eq. (1).

In a 3D FL, the width of the MDCs (Γ), or inverse of the QP scattering time ($\hbar/\tau$), is a quadratic function of binding energy. To check whether SrMnBi$_2$ can be described in such a framework, we analyzed the QP scattering rate by fitting the high-resolution MDCs along cut#1 and cut#2, as illustrated in Fig. 4a and Fig. 4b, respectively. Except below $E_B$=150 meV where the band splits and two Lorenztian curves are necessary to reproduce the results, a single Lorenztian curve is used to fit the MDCs up to 650 meV and 350 meV of binding energy along cut#1 and cut#2, respectively. We note that Lorentzians produce much better fit than Gaussians, suggesting that the spectral lineshape is at least not entirely resolution-limited. The values of the extracted QP MDC peakwidth as function of binding energy along the two cuts are shown in Figs. 4c and 4d. Considering the momentum resolution of the apparatus at the experimental setting, $\Delta k \sim 0.009$ Å at photon energy $h\nu=30$ eV, we remove its contribution from the fitted result using the formula $\Gamma_0 = \sqrt{\Gamma_{MDC}^2 - \Delta_k^2}$, and plot the intrinsic peakwidth in the corresponding figures. The intrinsic peakwidth in $k$ has $\Gamma_0 = \sqrt{\Gamma_{MDC}^2 - \Delta_k^2}$ also linear dependence with the binding energy.

The linear dependence of the linewidth is not expected by the FL theory, where moderate electron-electron interactions are the dominant term. The linear dependence of the scattering rate in a wide range is inconsistent with the 3D FL theory ($1/\tau \propto \omega^2$) or the 2D FL theory ($1/\tau \propto (\omega^2/\epsilon_F)ln(4\epsilon_F/\omega)$). In contrast, this kind of unusual behavior has been also observed in some other 2D systems such as graphene and layered electron gas[17,18], of which mechanism is not well understood. In graphene, it was either explained by the decay channel due to the formation of a plasmon band in layered electron gas, or by the strong electron-electron renormalization with similarity to the marginal Fermi Liquid (MFL) behavior[18]. It was also proposed that a pure graphene is a MFL whereas and doped graphene is a regular 2D FL, but most measurements have actually been performed on unintentionally doped graphene[3].

Although SrMnBi$_2$ is a 3D bulk crystal, our results suggest that the SrBi layer, which LDA calculations relate to the electronic states forming the Dirac cone, holds an unconventional 2D electron gas. Thus, SrMnBi$_2$ provides a clean system with undoped SrBi layers forming a 2D electron gas system for the study of low-dimensional systems, and calls for further theoretical and experimental studies.

**Conclusion**

In summary, we observed highly anisotropic Dirac-cones in bulk SrMnBi$_2$. Quasi-particles with linear dispersion are observed over a wide energy range. Their linewidth scattering rate increase linearly with binding energy, and the integrated density of states is also linearly dependent of binding energy. These behaviors suggests the existence of a novel two-dimensional electron gas in this three-dimensional bulk material.

**Materials and methods**

High quality single crystals of SrMnBi$_2$ were grown by the flux method. X-ray diffraction (XRD) data were collected with Cu $K_\alpha$ radiation at room temperature to confirm that the samples are from a single phase.. Samples were cleaved *in situ* yielding a flat (001) surfaces. ARPES measurements were performed at Renmin University of China using a Scienta R3000 analyzer with a He discharging lamp (He-I$\alpha$ line, $h\nu$ = 21.2 eV), at the PGM beamline of the Synchrotron Radiation Center (WI) using a Scienta R4000 analyzer, and at the SIS beamline at Swiss Light Source equipped with a R4000 analyzer. The high-resolution ARPES measurements were conducted with energy resolution better than 5 meV and angular resolution better than 0.2°, which corresponds to a momentum resolution of $\Delta k \leq 0.009$ Å$^{-1}$ at photon energy h$\nu$ = 30 eV.

**Figure Legends**

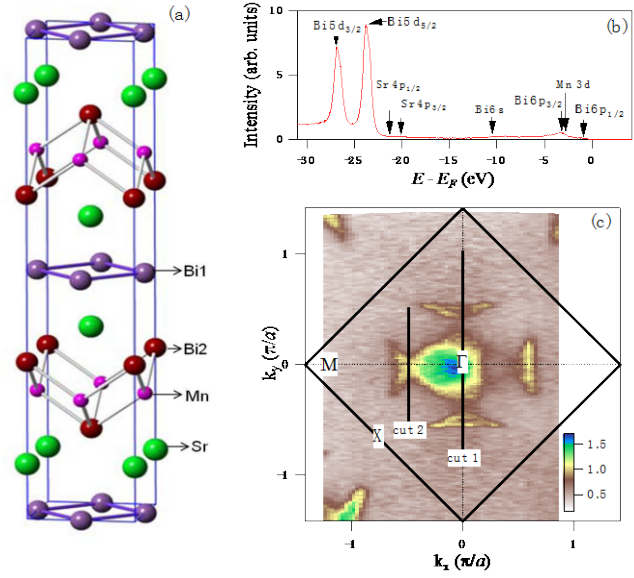

(**Figure 1**) **Structure, core levels and Fermi surface topology of SrMnBi$_2$.** (a) Crystal structure of SrMnBi$_2$ with MnBi$_4$ layers and Bi square nets. (b) Shallow core levels of the SrMnBi$_2$. (c) Integrated intensity plot around $E_F$ [-20meV, 20meV], which illustrates the FS topology of SrMnBi$_2$. The primary BZ is illustrated by black lines.

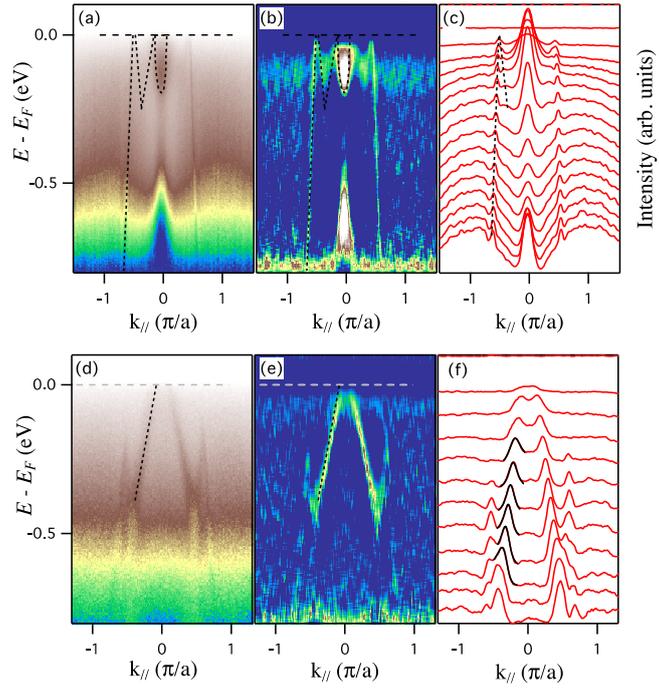

**(Figure 2) Band dispersion of anisotropic Dirac-cone along two directions.** (a)-(c) Intensity plot, second derivative $[\partial^2 I/\partial\omega^2]$, and EDCs along cut#1 shown in (c), respectively. (d), (e), (f) Intensity plot and EDC plot along cut#2.

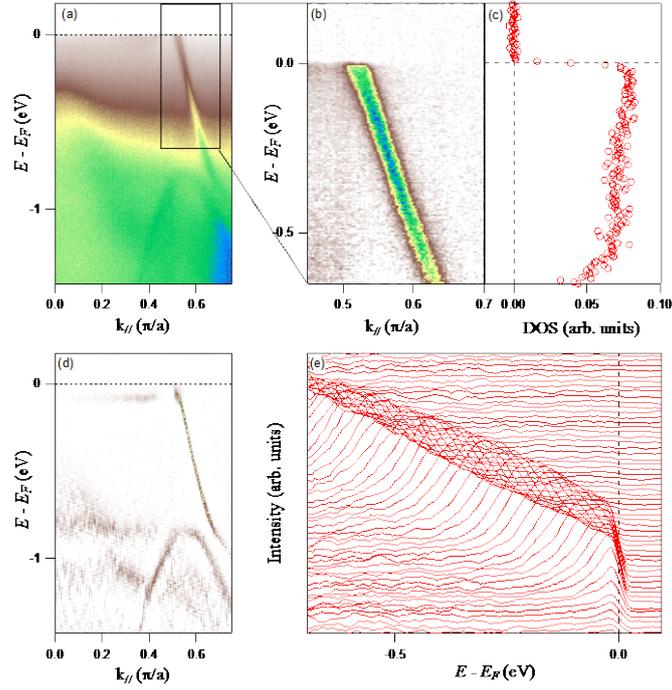

(**Figure 3**) **Constant line shape as function of binding energy** (a), (d) High resolution intensity plot and corresponding second derivative of the intensity along cut#1 direction up to 1.4 eV binding energy. (b), (e) Intensity plot and EDC plot of the band dispersion with background removed. (c) Integrated intensity of (b) as function of binding energy $n_k(\omega)$.

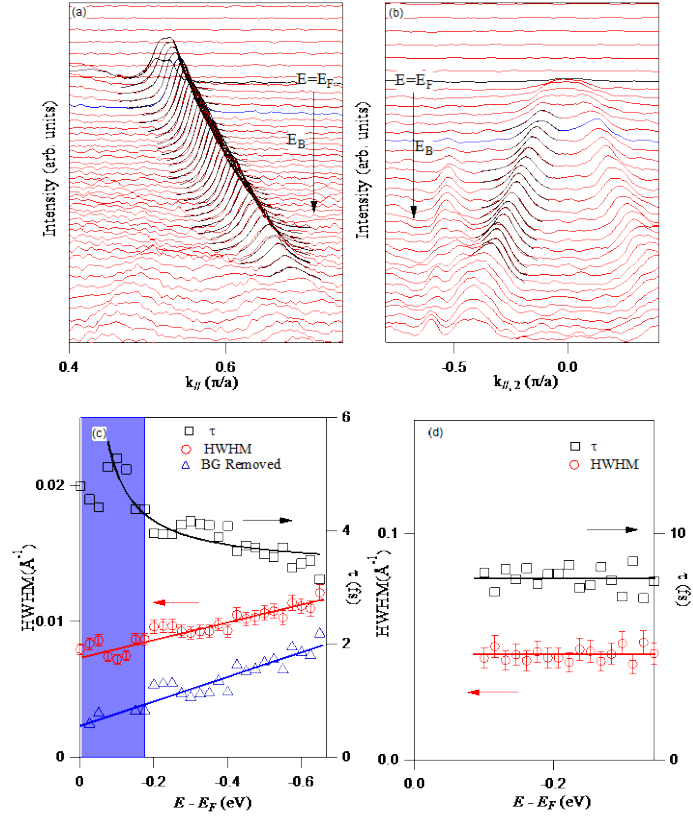

**(Figure 4) Unusual energy dependence of the quasi-particle width.** (a) MDCs and Lorentz-fits of the high-resolution intensity along cut#1, as shown in Fig.3b. (b) MDCs and Lorentz-fit results of the intensity plot along cut#2 direction. The black MDCs are recorded at $E_F$. The blue MDCs indicate the points between double-Lorentz and single-Lorentz fits. (c) and (d) Half width at half maximum (HWHM) (red circles) and lifetime (black squares) of the Dirac fermions along cut#1 and cut#2. The blue triangles in (c) show the intrinsic peakwidth, disposed of the momentum resolution. The lines in (c) and (d) are the fit results.

**Acknowledgements**

We thank for the kind experimental help from E. Rienks. This work is supported by grants from National Science Foundation of China, National Basic Research Program of China (973 Program), Ministry of Education of China, China Academic of Science and SSSTC. The Synchrotron Radiation Center, WI, is supported by University of Wisconsin-Madison with supplemental support from facility users and the University of Wisconsin-Milwaukee. This work is based in part upon research conducted at the Swiss Light Source, Paul Scherrer Institute, Villigen, Switzerland.

## Author contributions

**Competing Financial Interests statement**

The authors declare no competing financial interests.